\documentclass[12pt]{iopart}
\usepackage{graphicx}
\usepackage{amssymb}
\usepackage{dcolumn}
\usepackage{iopams}
\begin{document}

\title[Low carrier concentration crystals of the topological insulator Bi$_{2-x}$Sb$_{x}$Te$_{3-y}$Se$_{y}$]{Low carrier concentration crystals of the topological insulator Bi$_{2-x}$Sb$_{x}$Te$_{3-y}$Se$_{y}$:
a magnetotransport study}

\author{Y Pan$^1$, D Wu$^1$\footnote{Present address: School of Physics and Engineering, Sun Yat-Sen University, Guangzhou, China}, J R Angevaare$^1$, H Luigjes$^1$, E Frantzeskakis$^1$, N de Jong$^1$, E van Heumen$^1$, T V Bay$^1$, B Zwartsenberg$^1$, Y K Huang$^1$, M Snelder$^2$, A Brinkman$^2$, M S Golden$^1$ and A de Visser$^1$}

\address{$^1$ Van der Waals - Zeeman Institute, University of Amsterdam, Science Park 904, 1098 XH Amsterdam, The Netherlands}
\address{$^2$ Faculty of Science and Technology and MESA+ Institute for Nanotechnology, University of Twente, 7500 AE Enschede, The Netherlands}
\eads{\mailto{y.pan@uva.nl}, \mailto{a.devisser@uva.nl}}

\date{\today}

\begin{abstract} In 3D topological insulators achieving a genuine bulk-insulating state is an important research topic. Recently, the material system (Bi,Sb)$_{2}$(Te,Se)$_{3}$ (BSTS) has been proposed as a topological insulator with high resistivity and a low carrier concentration (Ren \textit{et al.} \cite{Ren2011}). Here we present a study to further refine the bulk-insulating properties of BSTS. We have synthesized Bi$_{2-x}$Sb${_x}$Te$_{3-y}$Se$_{y}$ single crystals with compositions around $x = 0.5$ and $y = 1.3$. Resistance and Hall effect measurements show high resistivity and record low bulk carrier density for the composition Bi$_{1.46}$Sb$_{0.54}$Te$_{1.7}$Se$_{1.3}$. The analysis of the resistance measured for crystals with different thicknesses within a parallel resistor model shows that the surface contribution to the electrical transport amounts to 97~\% when the sample thickness is reduced to $1~\mu$m. The magnetoconductance of exfoliated BSTS nanoflakes shows 2D weak antilocalization with $\alpha \simeq -1$ as expected for transport dominated by topological surface states.

\vspace{0.8cm}
\end{abstract}
\maketitle

\section{Introduction}

Three dimensional (3D) topological insulators (TIs) have generated intense research interest, because they offer unmatched opportunities for the realization of novel quantum states~\cite{Hasan&Kane2010,Qi&Zhang2010}. Theory predicts the interior of the TI sample (the bulk) is insulating, while the metallic surface states have a Dirac cone dispersion and a helical spin structure. Because of time reversal symmetry and a strong spin-orbit interaction, the surface charge carriers are insensitive to backscattering from non-magnetic impurities and disorder. This makes TIs promising materials for a variety of applications, ranging from spintronics and magnetoelectrics to quantum computation~\cite{Zutic2004,Qi2009,Nayak2008}. The topological surface states of exemplary TIs, such as Bi$_{1-x}$Sb$_{x}$, Bi$_{2}$Te$_{3}$, Bi$_{2}$Se$_{3}$, Sb$_{2}$Te$_{3}$, \textit{etc}., have been probed compellingly via surface-sensitive techniques, like angle-resolved photoemission spectroscopy (ARPES)~\cite{Hsieh2008,Xia2009,Chen2009,Hsieh2009b} and scanning tunneling microscopy (STM)~\cite{Roushan2009,Alpichshev2010}. However, the transport properties of the surface states turn out to be notoriously difficult to investigate, due to the dominant contribution from the bulk conduction resulting from intrinsic impurities and crystallographic defects. At the same time, potential applications strongly rely on the tunability and robustness of charge and spin transport at the device surface or interface. Therefore, achieving surface-dominated transport in the current families of TI materials remains a challenging task, in spite of the progress that has been made recently, including charge carrier doping~\cite{Analytis2010,Hor2009}, thin film engineering and electrostatic gating~\cite{Chen2010,Checkelsky2011}.

Recently, it was reported that the topological material Bi$_2$Te$_2$Se exhibits variable range hopping (VRH) behavior in the transport properties, which leads to high resistivity values exceeding $1~\Omega$cm at low temperatures~\cite{Ren2010}. This ensures the contribution from the bulk to electrical transport is small. At the same time the topological nature of the surface states was probed by Shubnikov - de Haas oscillations~\cite{Ren2010,Xiong2012}. Further optimizing studies include different crystal growth approaches~\cite{Jia2011}, Bi excess~\cite{Jia2012} and Sn doping~\cite{Ren2012,Fuccillo2013}. Bi$_2$Te$_2$Se has an ordered tetradymite structure (spacegroup R\={3}m) with quintuple-layer units of Te-Bi-Se-Bi-Te with the Te and Se atoms occupying distinct lattice sites. Next the composition of Bi$_2$Te$_2$Se was optimized by Ren \textit{et al.} \cite{Ren2011} by reducing the Te:Se ratio and introducing Sb on the Bi sites. An extended scan of isostructural Bi$_{2-x}$Sb$_{x}$Te$_{3-y}$Se$_{y}$ (BSTS) solid solutions resulted in special combinations of $x$ and $y$, notably $x=0.5$ and $y=1.3$, where the resistivity attains values as large as several $\Omega$cm at liquid helium temperature. In addition values of the bulk carrier concentration as low as 2$\times 10^{16}$~cm$^{-3}$ were achieved~\cite{Taskin2011}. The topological properties of BSTS samples with $x$ and $y$ near these optimized values were subsequently examined by a number of techniques, like ARPES~\cite{Arakane2012}, terahertz conductivity~\cite{Tang2013}, and STM and STS~\cite{Ko2013,Kim2014}.

In this paper we report an extensive study conducted to confirm and further investigate the bulk-insulating properties of Bi$_{2-x}$Sb$_{x}$Te$_{3-y}$Se$_{y}$ with $x$ and $y$ around 0.5 and 1.3, respectively. In the work of Ren \textit{et al.} \cite{Ren2011} the BSTS composition was scanned with a step size $\Delta x$ and $\Delta y$ of typically 0.25. For our study we prepared single crystals with much smaller step sizes, typically $\Delta x = 0.02$ and $\Delta y = 0.10$. Magnetotransport measurements showed that the carrier type in Bi$_{2-x}$Sb$_{x}$Te$_{1.7}$Se$_{1.3}$ changes from hole to electron when $x < 0.5$, while the carrier type remains unchanged in Bi$_{1.46}$Sb$_{0.54}$Te$_{3-y}$Se$_{y}$ when $y$ is varied from 1.2 to 1.6. The composition Bi$_{1.46}$Sb$_{0.54}$Te$_{1.7}$Se$_{1.3}$ presented the highest resistivity (12.6~$\Omega$cm) and lowest bulk carrier density ($0.2\times 10^{16}$~cm$^{-3}$) at $T = 4$~K.

In addition, the effect of reducing the sample thickness on the ratio between the surface and bulk conductivity was investigated. For this study we used the composition Bi$_{1.46}$Sb$_{0.54}$Te$_{1.4}$Se$_{1.6}$ and gradually thinned down a 140~$\mu$m thick sample to 6~$\mu$m. The analysis of the resistance data in terms of a two-resistor model reveals the surface contribution to the total conductivity can be as large as 97~\% at $T=4$~K when the sample thickness is $\sim$1~$\mu$m. The angular variation of the magnetoconductance of a nanoflake with composition Bi$_{1.46}$Sb$_{0.54}$Te$_{1.7}$Se$_{1.3}$ shows a pronounced weak antilocalization (WAL) term, whose field dependence followed the Hikami-Larkin-Nagaoka formula~\cite{Hikami1980} with a fit parameter $\alpha$ close to -1 as expected for topological surface states~\cite{Lu&Shen2011}. We conclude that good quality BSTS single crystals can be achieved via careful compositional variation and thickness reduction such that the topological surface transport overwhelms the bulk conduction channel.

\section{Methods}

High quality Bi$_{2-x}$Sb$_{x}$Te$_{3-y}$Se$_{y}$ single crystals were obtained by melting stoichiometric amounts of the high purity elements Bi (99.999~\%), Sb (99.9999~\%), Te (99.9999~\%) and Se (99.9995~\%). The raw materials were sealed in an evacuated quartz tube which was vertically placed in the uniform temperature zone of a box furnace. Both this choice of growth approach and our choice to keep the growth boules of small size (maximal dimension 1~cm) were made so as to maximise the homogeneity within each single crystal run. The molten material was kept at 850~$^{\circ}$C for 3~days and then cooled down to 520~$^{\circ}$C with a speed of 3~$^{\circ}$C/h. Next the batch was annealed at 520 $^{\circ}$C for 3 days, followed by cooling to room temperature at a speed of 10~$^{\circ}$C/min. In the following all $x$- and $y$-values refer to nominal concentrations. Electron probe microanalysis (EPMA) carried out on six crystals selected within the series showed the nominal compositions to be in good agreement with the stoichiometry in the final single crystals produced.  In addition, EPMA showed there to be no observable spatial inhomogeneity across the sample, which is in keeping with the homogeneous secondary electron images the crystals gave and the lack of any measurable impurity phases seen in standard x-ray diffraction characterisation of the samples. The systematic thickness dependence reported in the following also argues for an acceptable level of sample homogeneity. The as-grown platelet-like single crystals were cleaved with Scotch tape parallel to the $ab$-plane to obtain flat and shiny surfaces at both sides. Care was taken to maintain a sample thickness of $\sim 100~\mu$m for all samples. Next the samples were cut into a rectangular shape using a scalpel blade. For the transport measurements in a five-probe configuration, thin (50~$\mu$m) copper wires were attached to the samples using silver paint. Current and voltage contacts were made at the edges of the sample to ensure contact with the bulk and the upper and lower surface. The exposure time to air between cleaving and mounting the samples in the cryostat was kept to a minimum of about 1~h.

Four-point, low-frequency, ac-resistivity and Hall effect measurements were performed in a MaglabExa system (Oxford Instruments) equipped with a 9~T superconducting magnet in the temperature range from 4 to 300~K. The excitation current was typically 1000~$\mu$A. Selected measurements were spot-checked using a PPMS system (Dynacool, Quantum Design) with 100~$\mu$A excitation current. Measurements were always made for two polarities of the magnetic field after which the Hall resistance, $R_{\rm {xy}}$, and longitudinal resistance, $R_{\rm {xx}}$, were extracted by symmetrization. When investigating the effect of the sample thickness, layers were removed from the sample by Scotch tape. Special care was taken to maintain a uniform thickness across the sample, as well as the same lateral dimensions. For the thickness-dependent series, the resistance measurements were made in a bath cryostat in the temperature range 4.2-300~K using an AC Resistance Bridge (Model 370, LakeShore Cryotronics).

For our investigation of the WAL effect, flakes of thickness in the range of 80 to 200~nm were mechanically exfoliated on silicon-on-insulator substrates. Au Hall bar electrodes were prepared by standard photolithography followed by e-beam lithography and argon ion etching to shape the flake in a Hall bar structure. During the fabrication steps we covered the devices with e-beam or photoresist to avoid damaging and contamination of the surface. The Hall bars have a total length of 24~$\mu$m and have widths in the range of 0.5-2.0~$\mu$m. Transport measurements on these samples were carried out in a PPMS (Dynacool, Quantum Design) in the temperature range 2-300~K with an excitation current of 5~$\mu$A. The field-angle dependence of WAL was measured for a rotation of the Hall bar around its long axis (the current direction).

ARPES measurements were carried out on cleaved single crystals of Bi$_2$Se$_3$ and Bi$_{1.46}$Sb$_{0.54}$Te$_{1.7}$Se$_{1.3}$ at the SIS beamline of the Swiss Light Source at the Paul Scherrer Institute. The photon energies used were 27 and 30~eV, and the sample temperature was 17~K. In both cases, the surfaces were exposed to sufficient residual gas such that the downward band bending -- as documented in \cite{King2011,Bahramy2012} for Bi$_2$Se$_3$ -- has saturated and thus was no longer changing as a function of time. The energy resolution was 15~meV.

\section{Results and analysis}

\subsection{Composition variation}

\begin{figure}
\begin{center}
\includegraphics[height=11cm]{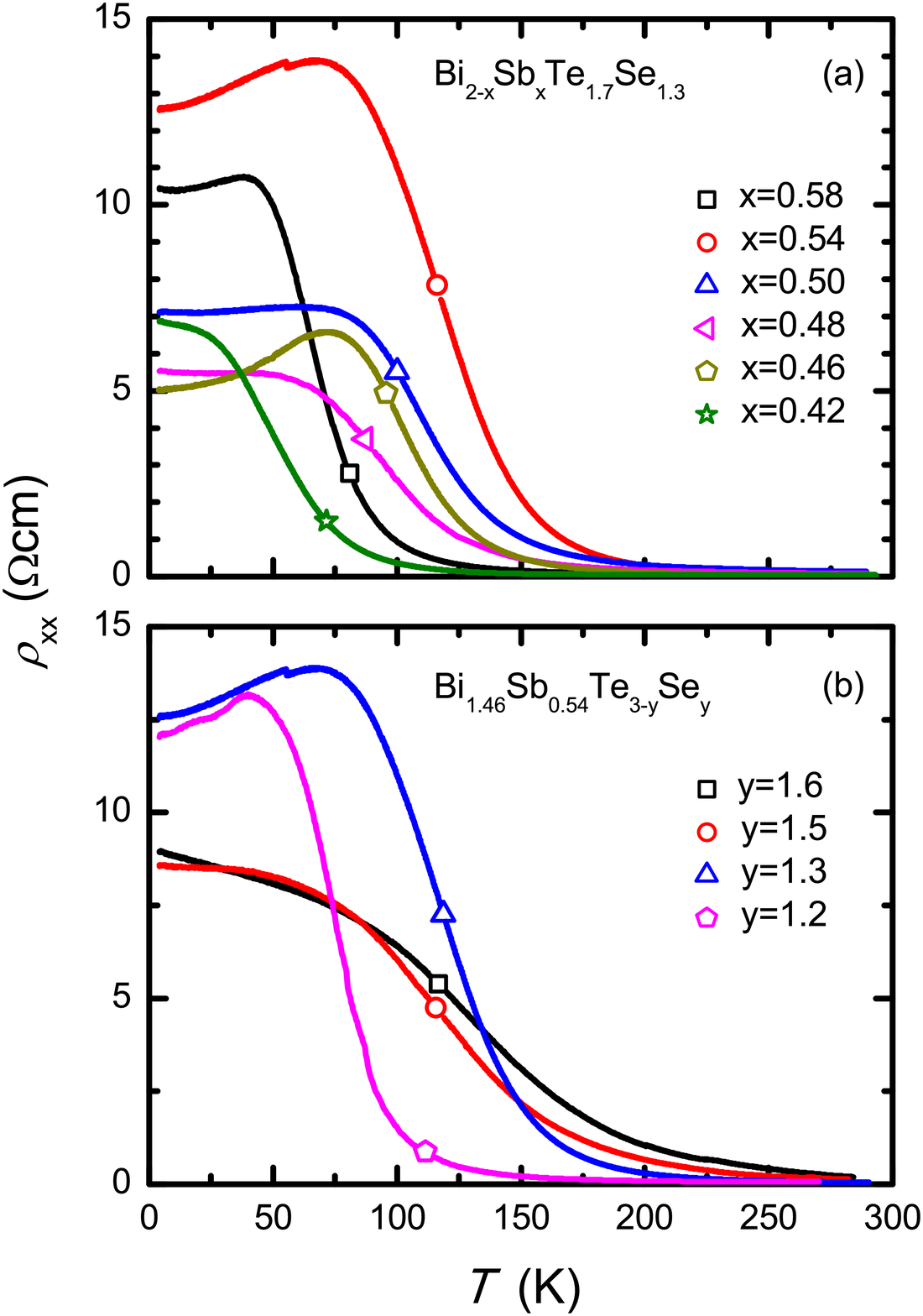}
\caption{Temperature dependence of the resistivity of Bi$_{2-x}$Sb$_{x}$Te$_{3-y}$Se$_{y}$ crystals with (a) $y=1.3$ and $x$-values as indicated, and (b) $x=0.54$ and $y$-values as indicated. The typical sample thickness is $100$~$\mu$m.}
\end{center}
\end{figure}

\begin{figure}
\begin{center}
\includegraphics[height=12cm]{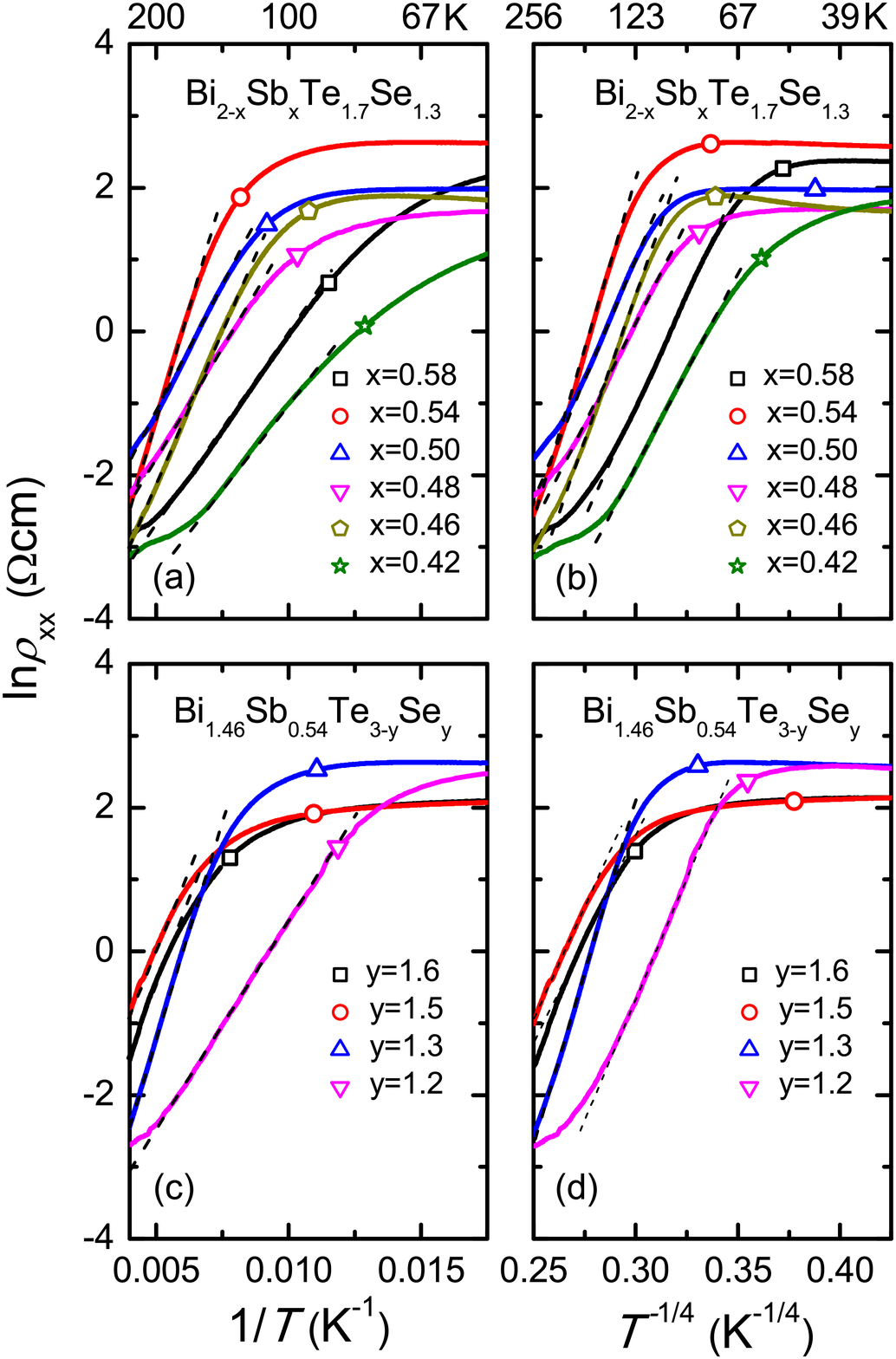}
\caption{Temperature variation of the resistivity of BSTS in a plot of $\ln \rho_{\rm {xx}}$ \textit{versus}  $1/T$ in frames (a) and (c), and \textit{versus} $T^{-1/4}$ in frames (b) and (d). The linear dashed lines represent the activation behavior and the 3D VRH behavior (see text). In frames (a) and (b) we have traced the resistivity for $y=1.3$ and $x$-values as indicated, and in frames (c) and (d) the resistivity for $x=0.54$ and $y$-values as indicated.}
\end{center}
\end{figure}

In this section we present the resistivity, $\rho_{\rm {xx}}$, and Hall effect, $\rho_{\rm {xy}}$, data. Rather than simply re-growing the optimal $(x,y)$-values of (0.5,1.3) reported by Ren \textit{et al.} \cite{Ren2011}, we first varied $x$ from 0.42 to 0.58 while keeping $y$ fixed at 1.3. Next, we fixed $x$ at 0.54 and varied $y$ from 1.2 to 1.6. The size of the steps taken in both $x$ and $y$ were smaller than those reported earlier. The temperature variation of the resistivity for these two series of crystals is shown in Figure~1a and Figure~1b, respectively. All the samples display an overall similar resistivity behavior. Upon cooling below 300~K the resistivity increases first slowly and then faster below $\sim 150$~K till $\rho_{\rm {xx}}$ reaches a maximum value at 50-100~K. Next the resistivity shows a weak decrease and levels off towards 4~K. The increase of the resistivity can be described by an activated behavior $\rho _{\rm {xx}} \propto \exp (\Delta /k_{\rm B} T)$ where $\Delta$ is the activation energy, followed by a 3D VRH regime $\rho _{\rm {xx}} \propto \exp [(T/T_0)^{-1/4}])$ where $T_0$ is a constant~\cite{Ren2011}. These different regimes show considerable overlap as illustrated in Figure~2a and 2b, and in Figure~2c and 2d. Below $\sim 50$~K the resistivity is described by a parallel circuit of the insulating bulk and the metallic surface states (see the next section).

\fulltable{\label{tab:table1} Transport parameters obtained from resistivity and Hall effect measurements of BSTS crystals with composition as given in the first column: $\rho_{\rm {xx}}(280$~K), $\rho_{\rm {xx}}(4$~K), activation energy $\Delta$ estimated from the linear part in $\ln \rho_{\rm {xx}}$ \textit{vs.} $1/T$, the low-field Hall coefficient $R_{\rm {H}}(200$~K) and $R_{\rm {H}}(4$~K), the bulk carrier density $n_{\rm {b}}$ and the transport mobility $\mu$ calculated in a single band model.}
\br
 Composition & $\rho_{\rm {xx}}(280$~K$)$ & $\rho_{\rm {xx}}(4$~K$)$ & $~~\Delta$ & $R_{\rm H}(200$~K$)$ & $R_{\rm H} (4$~K$)$ & $n_{\rm b}(4$~K$)$ &$\mu(4$~K$)$  \cr
Bi$_{2-x}$Sb$_{x}$Te$_{3-y}$Se$_{y}$ & (m$\Omega$cm) &($\Omega$cm) & (meV) & (cm$^3$/C) & (cm$^3$/C) & (10$^{16}$~cm$^{-3}$) & (cm$^2$/Vs) \cr

\mr
  $x=0.58$ ; $y=1.3$ & \050.5 & 10.4 & \044& \m--- & \,\0-468 & \0\01.3 & \045 \cr
  $x=0.54$ ; $y=1.3$ & \053.6 & 12.6 & 102 & \030 & \,-3110 & \0\00.2 & 247 \cr
  $x=0.50$ ; $y=1.3$ & 141 & \07.1 & \056 & \027 & \0\0380 & \0\01.6 & \054 \cr
  $x=0.48$ ; $y=1.3$ & \081.9 & \05.5 & \048 & \020 & \,\0-167 & \0\03.7 & \030 \cr
  $x=0.46$ ; $y=1.3$ & \036.6 & \05.0 & \072 & \031 & \01657 & \0\00.4 & 331 \cr
  $x=0.42$ ; $y=1.3$ & \044.4 & \06.8 & \036 & \0\04 & \0\0\0\06 & 107 & \0\00.9 \cr
  $x=0.54$ ; $y=1.6$ & 133 & \08.6 & \083 & \,-92 & \,\0-707 & \0\00.9 & \082 \cr
  $x=0.54$ ; $y=1.5$ & 214 & \08.9 & \066 & \037 & \,\0-278 & \0\02.2 & \031 \cr
  $x=0.54$ ; $y=1.2$ & \m--- & 12.1 & \048 & \0\08 & \,-1940 & \0\00.3 & 160 \cr
\br
\endfulltable

In Table~1 we have collected the $\rho_{\rm {xx}}$-values at 280~K and 4~K and the activation energy $\Delta$ for a number of samples. The data in this table represent the summary of 35 individual measurements, 15 of which were carried out on the $x=0.54$ and $y=1.3$ composition. The resistivity values at $T = 4$~K all exceed $5~\Omega$cm. We remark that in our first series of crystals with a fixed value $y=1.3$ the sample with $x=0.54$ has a record-high $\rho_{\rm {xx}}$-value of $12.6~\Omega$cm at $T = 4$~K, and the largest value $\Delta = 102$~meV as well. Subsequent small changes of $y$ in the range 1.2-1.6 did not yield a further increase of the low temperature resistivity.

\begin{figure}
\begin{center}
\includegraphics[height=11cm]{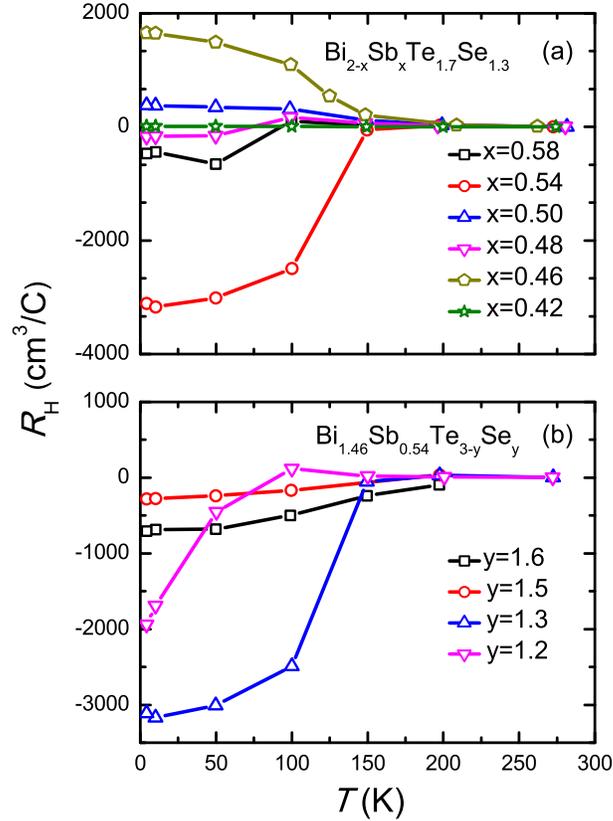}
\caption{Temperature dependence of the low-field Hall coefficient $R_{\rm {H}}$ of BSTS crystals with (a) $y=1.3$ and $x$-values as indicated, and (b) $x=0.54$ and $y$-values as indicated.}
\end{center}
\end{figure}

Given these resistivity data, Hall experiments are of interest to examine the character and quantity of bulk carriers in these samples. The temperature variation of the low-field Hall coefficient $R_{\rm H}(T)$ for the two series of crystals with fixed $y=1.3$ and fixed $x=0.54$ is shown in Figure~3a and 3b, respectively. The Hall coefficients were obtained by fitting the linear Hall resistivity for $B \leq 1$~T.   For the first series (Figure~3a), $R_{\rm H}$ of all samples is positive at temperatures above 200 K and the values gradually increase upon lowering the temperature. Near 150~K, \textit{i.e.} near the temperature where the resistivity starts to rise quickly, the absolute value of the Hall coefficient increases rapidly. For an increasing Sb content with respect to $x \approx 0.5$, $R_{\rm H}$ turns negative, whereas for a decreasing Sb content $R_{\rm H}$ remains positive. For the second series of samples (Figure~3b) $R_{\rm H}$ of all samples starts positive near room temperature, but eventually attains fairly large negative values near 4~K.

The carrier type and concentration in BSTS and related compounds is in general connected to the competition between two effects: (\textit{i}) (Bi,Sb)/Te antisite defects which act as electron acceptors (or hole dopants), and (\textit{ii}) Se vacancies which act as electron donors. This tells us that in the high temperature regime where $R_{\rm H} > 0$ (Bi,Sb)/Te antisite defects are dominant in providing carriers. However, at low temperatures $R_{\rm H} < 0$ and Se vacancies prevail, except for the crystals with a reduced Sb content ($x < 0.5$). $R_{\rm H}$-values at 200~K and 4~K are listed in Table 1. In order to compare the transport parameters with values reported in the literature we have listed the bulk carrier density, $n_{\rm {b}}$, and mobility $\mu = 1/ne \rho_{\rm {xx}}$ as well, assuming a simple single band model. Clearly, in these data, the crystal with the composition Bi$_{1.46}$Sb$_{0.54}$Te$_{1.7}$Se$_{1.3}$ stands out as the one with the highest resistivity, the largest activation gap and the lowest carrier concentration $n_b = 0.2\times 10^{16}$~cm$^{-3}$.

We grew three batches of single crystals of this composition and the measurement of 15 single crystals of Bi$_{1.46}$Sb$_{0.54}$Te$_{1.7}$Se$_{1.3}$ gave a reproducible picture that this composition generally gave the most bulk-insulating behaviour. There is some sample to sample variation in the transport parameters. For example, the low-T resistivity for the $x=0.54$/$y=1.3$ composition of 12.6~$\Omega$cm given in Table 1 was representative, but values were also measured to be in the range of 11-15~$\Omega$cm.

\subsection{Thickness variation}

\begin{figure}
\begin{center}
\includegraphics[height=11cm]{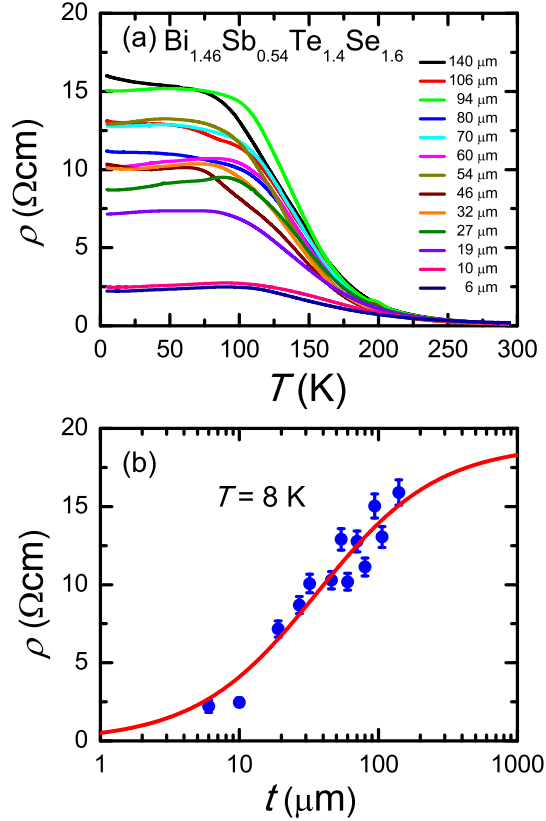}
\caption{(a) Temperature dependence of the resistivity of a Bi$_{1.46}$Sb$_{0.54}$Te$_{1.4}$Se$_{1.6}$ crystal with thickness $t= 140~\mu$m thinned down in 12 steps to 6~$\mu$m as indicated. (b) Measured resistivity at $8$~K as a function of thickness (solid dots). The solid line represent a fit of the data to the parallel resistor model (see text). }
\end{center}
\end{figure}

A simple and elegant way to separate the contribution from the surface and the bulk to the total conductivity of a topological insulator is by reducing the sample thickness, $t$~\cite{Taskin2011,Xia2013,Kim2014a,Syers2014}. In Figure~4a we show the resistivity $\rho (T)$ of a Bi$_{1.46}$Sb$_{0.54}$Te$_{1.4}$Se$_{1.6}$ crystal with $t= 140~\mu$m that subsequently was thinned down in 12 steps to 6~$\mu$m. Here $\rho = (A/l) \times R$ where $A=t \times w$ is the cross sectional area for the current ($w$ is the sample width) and $l$ the distance between the voltage contacts. The overall behavior $\rho (T)$ for all thicknesses is similar to the results presented in Figure~1. However, while the curves almost overlap at high temperatures, the $\rho$-value at low temperatures decreases significantly when reducing the sample thickness. This tells us the ratio between the surface and bulk contribution changes with thickness.

For a proper analysis the parallel resistor model is used:
\begin{equation}
\rho= \frac{\rho_{\rm b} \rho_{\rm s} (t_{\rm b} + 2 t_{\rm s})}{2 t_{\rm s} \rho_{\rm b} + t_{\rm b} \rho_{\rm s}}
\end{equation}
where $\rho_{\rm b}$ and $\rho_{\rm s}$ are the resistivities, and $t_{\rm b}$ and $t_{\rm s}$ the thicknesses, of the bulk and the surface layer~\cite{Ren2011}, respectively. For $t_{\rm s}$ we may take 3~nm which is the thickness of 1 unit cell. The factor 2 in the equation above counts the top \textit{and} the bottom surface of the sample. In Figure~4b we have traced $\rho$ taken at 8~K as a function of $t \cong t_{\rm b}$. The uncertainty in the $\rho$-values is mainly due to the error in the geometrical factor, especially in the value of $l$ ($\pm 5$\%) because of the finite size of the silver paint contacts. The solid line represents a least square fit to Equation~1 with fit parameters $\rho_{\rm b} = 19.0~\Omega$cm and $\rho_{\rm s} = 3.14 \times 10^{-3}~\Omega$cm, or expressed as conductivities $\sigma_{\rm b} = 0.053~\Omega^{-1}$cm$^{-1}$ and $\sigma_{\rm s} = 318~\Omega^{-1}$cm$^{-1}$. Consequently, the surface conductance $G_{\rm s} = 2 \times t_{\rm s} \times \sigma_{\rm s} = 1.91 \times 10^{-4}~\Omega^{-1}$. We remark our value for $\sigma _{\rm b}$ compares favorably to (is smaller than) the values 0.1~$\Omega^{-1}$cm$^{-1}$ and 0.12~$\Omega^{-1}$cm$^{-1}$  for sample compositions Bi$_{1.5}$Sb$_{0.5}$Te$_{1.7}$Se$_{1.3}$ and Bi$_{1.5}$Sb$_{0.5}$Te$_{1.8}$Se$_{1.2}$ reported in \cite{Taskin2011} and \cite{Xia2013}, respectively. The ratio of the surface conductance over the total sample conductance can be calculated as $G_{\rm s} /(G_{\rm s} + t \sigma_{\rm b})$. With our fit parameters we calculate for samples with a thickness of 100, 10 and 1~$\mu$m a surface contribution of 27~\%, 78~\% and 97~\%, respectively. For a nanoflake with typical thickness of 130 nm (see the next section), we obtain a value of 99.6~\%. We conclude surface-dominated transport can be achieved in our BSTS crystals grown with a global composition Bi$_{1.46}$Sb$_{0.54}$Te$_{1.4}$Se$_{1.6}$ when the sample thickness is less than $\sim 1~\mu$m.

\subsection{Weak antilocalization}

The thickness dependence of the resistivity shows dominance of surface transport for thin samples. A further test as to whether the surface dominated transport is consistent with the presence of topological surface conduction channels is to search for and characterise possible signals of weak antilocalization. For our study of weak antilocalization we selected the BSTS composition we found to give crystals with the highest bulk resistivity: Bi$_{1.46}$Sb$_{0.54}$Te$_{1.7}$Se$_{1.3}$. The magnetoresistance of an exfoliated nanoflake, structured by e-beam lithography into a Hall bar, was measured in the temperature range 2-40~K and in magnetic fields up to 2~T. The dimensions of the Hall bar are: thickness $t= 130 \pm 5~$nm, channel width $w =2 \pm 0.02~\mu$m and distance between the voltage contacts $l =6.75 \pm 0.25~\mu$m. The error in $l$ takes into account the extended size of the voltage electrodes. The temperature variation of the resistivity $\rho(T)$ of the Hall bar is shown in the inset of Figure~5b. For this thickness the resistivity at low temperatures levels off at a low value of $0.035~\Omega$cm, in good agreement with the functional behavior reported in Figure~4b for BSTS with a Se content $y=1.6$. Since $\rho_{\rm b} \gg \rho_{\rm s}$ we obtain a sheet or surface conductance $G_{\rm s} \approx t / \rho = 3.7 \times 10^{-4} ~\Omega ^{-1}$.

\begin{figure}
\begin{center}
\includegraphics[height=11cm]{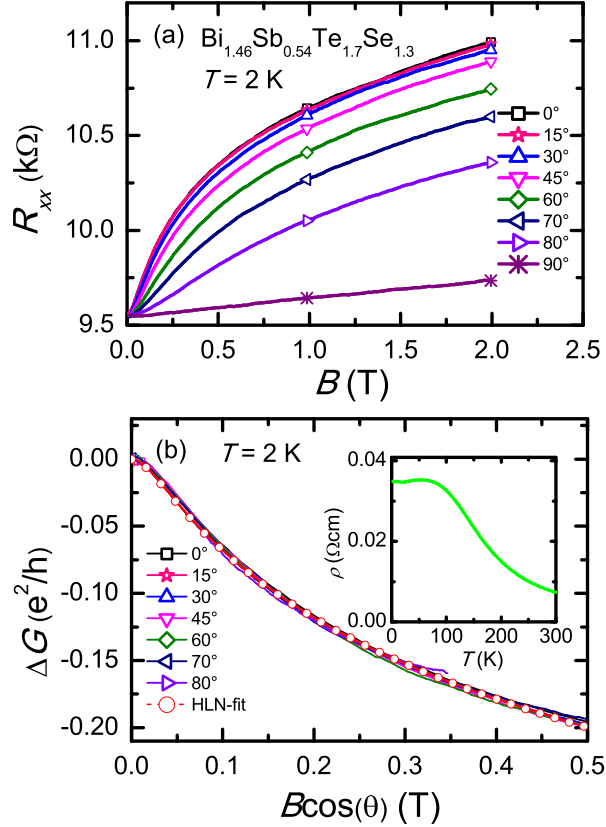}
\caption{(a) Longitudinal resistance $R_{\rm {xx}}$ of a BSTS Hall bar fabricated from a 130~nm thick flake measured at $T=2$~K as a function of the magnetic field for different field angles as indicated. For $\theta = 0^{\circ}$, $B$ is directed perpendicular to the sample surface. (b) Magnetoconductance $\Delta G$ at $T=2$~K plotted as a function of $B \cos \theta$, where $\theta$ is the angle between the surface normal and $B$. The data collapse onto an universal curve, which confirms the 2D nature of WAL. The open circles represent a fit to the HLN expression (Equation~2) in the field range $B_{\perp} = 0 - 0.5$~T. The fit parameters are $l_\phi = 116$~nm and $\alpha = -1.14$. Inset: $\rho_{\rm {xx}}$ of the BSTS Hall bar as a function of temperature. }
\end{center}
\end{figure}

\begin{figure}
\begin{center}
\includegraphics[height=6cm]{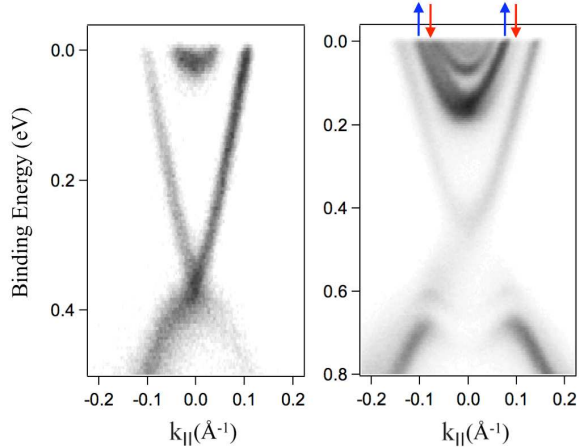}
\caption{ARPES data recorded from the surfaces of Bi$_{1.46}$Sb$_{0.54}$Te$_{1.7}$Se$_{1.3}$ (left) and Bi$_2$Se$_3$ (right) under conditions of saturated downward band bending due to adsorption of residual gas atoms on the surface. The data cut through the center of the surface Brillouin zone were measured using photon energies of 27~eV (BSTS) and 30~eV (Bi$_2$Se$_3$) at a temperature of 17~K. The arrows in the right panel denote the spin polarization of the Rashba-type surface states observed on band-bent  Bi$_2$Se$_3$. Notice the different energy scales of the two panels}
\end{center}
\end{figure}

In Figure~5a we show the resistance, $R_{\rm {xx}}$, as a function of $B$, measured at $T=2$~K. The relatively sharp increase of $R_{\rm {xx}}$ with field is ubiquitous in TIs~\cite{Culcer2012,Lu&Shen2011,Lu&Shen2014} and is attributed to the suppression of 2D weak antilocalization, \textit{i.e.} the constructive interference of time reversed scattering loops, generated by a magnetic field applied perpendicular to the sample surface, $B_{\perp}$. As the analysis of the magnetoresistance data we carry out below will show, the phase coherence length $l_{\rm {\phi}}$ for the WAL in this system is of the same order as the flake thickness, justifying a double-check that the WAL behavior is indeed two-dimensional  by recording the angular dependence of the WAL. In Figure~5a, the magnetoresistance, $\Delta R \equiv R(B)-R(0)$ (we have dropped the subscript $\rm {xx}$) is seen to be largest for $B$ perpendicular to the sample surface ($\theta = 0 ^{\circ}$). For an in-plane magnetic field ($\theta = 90 ^{\circ}$) a small residual magnetoresistance is seen of unknown origin, but this has essentially no effect on the parameters coming out of the fits to the data presented and discussed in the following. In Figure~5b we trace the magnetoconductance $\Delta G_{\rm {2D}}(B,\theta) \equiv  G(B,\theta) - G(B,\theta =90 ^{\circ})$ for $0 ^{\circ}  \le \theta \le 80 ^{\circ}$ as a function of $B \cos \theta$. Obviously, the data collapse onto an universal curve, which signals the two dimensional character of the WAL in these samples.

\fulltable{\label{tab:table2} Phase coherence length $l_{\phi}$ and prefactor $\alpha$, obtained by fitting the 2D magnetoconductance of BSTS crystals to the HLN expression, Equation 2. $B_{\rm {\perp , max}}$ gives the field range for the fit (see text). }
\br
 Composition & Thickness & $B_{\rm {\perp ,max}}$ & $l_{\rm {\phi}}$ & $\alpha$ & Temperature & Reference  \\
Bi$_{2-x}$Sb$_{x}$Te$_{3-y}$Se$_{y}$ & (nm) &(T) &(nm) & &(K) & \\
\mr
  $x=0.54$ ; $y=1.3$ & 130 & 0.5 &  116 & -1.14 & \02 &this work \\
  $x=0.54$ ; $y=1.3$ & 130 & 0.5 &  \052 & -0.93   & 40 &this work \\
  $x=0.5~~$ ; $y=1.3$ & 200 & 0.5 &  121 & -0.96 & \02 &\cite{Hsiung2013} \\
  $x=0.5~~$ ; $y=1.2$ & 596 & 0.5 &  160 & -0.75 & \02 &\cite{Xia2013}  \\
  $x=0.5~~$ ; $y=1.3$ & \085 & 6 &  170 & -1.3 & \04 & \cite{Lee2012} \\
\br
\endfulltable

Next we compare the universal curve with the expression for the magnetoconductivity $\Delta \sigma _{\rm {2D}}$ of 2D weak antilocalization put forward by Hikami, Larkin and Nagaoka (HLN)~\cite{Hikami1980}:

\begin{equation}
\Delta \sigma_{\rm {2D}}= -\frac{\alpha e^2}{2\pi^2 \hbar}\left(\ln\left(\frac{\hbar}{4 e l_{\rm {\phi}}^2 B}\right)-\psi\left(\frac{1}{2}+\frac{\hbar}{4 e l_{\rm {\phi}}^2 B}\right)\right)
\end{equation}

Here $\psi$ is the digamma function, $l_{\rm {\phi}}$ is the effective phase coherence length and $\alpha$ is a prefactor. Each scattering channel from a band that carries a $\pi$ Berry phase contributes a value $\alpha = - 1/2$~\cite{Lu&Shen2011}. For independent bottom and top topological surface channels we therefore expect $\alpha = - 1$. In Equation 2, $\Delta \sigma_{2D} = \frac{l}{w} \Delta G_{2D}$, where $\frac{l}{w} = 3.38 \pm 0.13$. We have fitted the collapsed magnetoconductance data at $T=2$~K to the HLN expression with $l_{\rm {\phi}}$ and $\alpha$ as fit parameters. In the field interval $B \cos \theta = 0-0.5$~T we find $l_{\rm {\phi}} = 116$~nm and $\alpha = -1.14 \pm 0.06$. For this value of $l_{\rm {\phi}}$ we calculate $B_{\phi} = \hbar/4el_{\rm {\phi}}^2 =0.012$~T and the condition $B \gg B_{\phi}$ is easily met, \textit{i.e.} the applied field is sufficiently large to suppress WAL. Furthermore, $l_{\rm {\phi}} \ll w$ which ensures our Hall bar has the proper dimensions for 2D WAL. If we extend the field range of the fit, the value of $\alpha$ decreases slightly, but at the same time the quality of the fit decreases. For instance, by fitting up to 2~T we obtain $\alpha = -0.96 \pm 0.06$.

The temperature variation of the field-induced suppression of WAL has been investigated at $T= 2, 4, 10, 15, 25$ and $40~$K. By increasing the temperature the WAL becomes weaker and is suppressed more easily by the magnetic field. By fitting our collapsed $\Delta G_{\rm {2D}} (B_{\perp})$ curves, $l_{\rm {\phi}}$ decreases to 52~nm at $T= 40$~K. A power-law fit yields $l_{\rm {\phi}} \propto T^{-0.46}$ in the temperature range 10-40~K, which is close to the expected behavior $T^{-0.5}$ for 2D WAL. However, below 10~K $l_{\rm {\phi}}$ levels off and the exponent drops to -0.06. The prefactor $\alpha$ decreases slightly to $-0.93 \pm 0.05$ at 40~K for a fit range $0-0.5$~T.

\section{Discussion}

Single crystals of Bi$_{2-x}$Sb$_{x}$Te$_{3-y}$Se$_{y}$ with $x$ and $y$ around 0.5 and 1.3 can nowadays clearly be considered to have superior bulk insulating properties~\cite{Ren2011,Taskin2011,Xia2013,Hsiung2013,Lee2012} amongst the 3D topological insulators. This makes BSTS particularly attractive for exploratory research into TI devices with functionalities based on protected surface transport. In this paper we have shown that by fine tuning the BSTS composition it is possible to obtain record high values of the resistivity. For a global composition of Bi$_{1.46}$Sb$_{0.54}$Te$_{1.7}$Se$_{1.3}$ low-T resistivity values exceeding $10~\Omega$cm were common for samples with a thickness of 100~$\mu$m. At the same time Hall data show the bulk carrier concentration in these samples can be as low as $0.2\times 10^{16}$~cm$^{-3}$ at $T = 4$~K. The strong bulk insulating behavior is furthermore demonstrated by the large value of the activation gap for transport, $\Delta \simeq 100$~meV. The data of Table 1 show there to be a non-trivial dependence of the transport characteristics on the $x$ and $y$-values chosen for each single crystal batch. Our sample characterization shows that this is clearly not a consequence of random compositional irreproducibility from sample to sample. Rather, the chemistry controlling the defect density and balance (between p- and n-type) in these low-carrier concentration 3D TI materials is a subtle quantity, and not one that simply tracks the Bi/Sb and Te/Se ratios in a linear fashion. This makes a variation of composition necessary to determine the best route to the most bulk insulating behaviour. This was done in Ref.~\cite{Ren2011} and is also the approach adopted here, resulting in excellent bulk-insulating characteristics.

Because of the genuine bulk insulating behavior of the optimized BSTS crystals, the $\rho$-values obtained at low temperatures are not true resistivity values but depend on the thickness when $t \leq 1$~mm (see Figure~4). Therefore the transport data result from parallel channels due to the bulk and top/bottom surfaces. Our analysis with the parallel resistor model of the resistance measured on the very same BSTS crystal made thinner and thinner shows that the surface contribution to the total transport is close to 97~\% for a thickness of $1~\mu$m. Hence we conclude that devices fabricated with submicrometer thickness are sufficiently bulk insulating to exploit the topological surface states by transport techniques. The nanoflake with a thickness of 130~nm certainly fulfills this condition.

The HLN fit of the magnetoconductance of the thin flake yields values of $\alpha$ close to -1, which is in keeping with the thickness dependence of the resistivity shown in Figure~4 in that for a flake of this thickness essentially all of the transport occurs through the surface states. The $\alpha$ value of close to -1 would suggest that it is the topological surface states at the top and bottom of the crystal which are  dominating the transport. In Table 2 we have collected the fit parameters $l_{\rm {\phi}}$ and $\alpha$, as well as those obtained for BSTS nanoflakes reported in the recent literature. Hsiung \textit{et al.} \cite{Hsiung2013} measured a flake with enhanced surface mobility and reported an $\alpha$-value close to -1 as well. The $\alpha$-values reported by Xia \textit{et al.} \cite{Xia2013} obtained on a relatively thick nanoflake are systematically smaller than -1. Lee \textit{et al.} \cite{Lee2012} found a significantly larger value $\alpha = -1.3$ for a gated nanoflake at zero bias. Here the large value of $\alpha$ is attributed to the combined presence of non-trivial and Rashba-split conduction channels in the topologically trivial 2DEG caused by band bending.

The existence (or not) of Rashba spin-split states at the surface of BSTS is of importance for the magnetoconductance -- and in particular -- the WAL behaviour. The alpha value we extract for BSTS of close to -1 (Table 2) suggests there is not a contribution from Rashba spin-split states in our BSTS crystals. In Bi$_2$Se$_3$ it is well established from ARPES that topologically trivial, confined bulk states can form a 2DEG in the near surface region and that these states can also show Rashba-type spin splitting~\cite{King2011,Bahramy2012}. Thus, in Figure 6 we show two ARPES images of the portion of $k$-space near the Gamma-point: one for Bi$_2$Se$_3$ and one for a BSTS crystal from the same source and of the same composition as the flake used for the WAL studies. In both cases, we show ARPES data from surfaces after significant exposure to adsorbates - \textit{i.e.} for the case of essentially maximal band-bending. The Bi$_2$Se$_3$ data show clear signs of confinement of states related to both the conduction and valence bands, as well as of emerging Rashba-type spin splitting for the states crossing the Fermi level, in accordance with the literature~\cite{King2011,Bahramy2012}. The left panel of Figure 6 shows analogous data for adsorbate exposed BSTS. The downward band bending has led to the population of states derived from the bulk conduction band, but there is clearly no Rashba-type spin splitting at the BSTS surface. This is fully consistent with the alpha value close to -1 extracted from the analysis of the WAL data shown in Figure~5 and Table 2.

\section{Summary}

We have presented an extensive study of the bulk-insulating properties of BSTS single crystals. We have synthesized numerous Bi$_{2-x}$Sb${_x}$Te$_{3-y}$Se$_{y}$ single crystals with compositions around $x = 0.5$ and $y = 1.3$, with steps in $x$ of 0.02 and $y$ of 0.1. The samples were investigated by resistance and Hall effect measurements. We show that via variation of the composition on a fine level, we could arrive at a record-high resistivity, bulk-insulating transport behaviour and a low carrier density, \textit{e.g.} for multiple growth runs for the composition Bi$_{1.46}$Sb$_{0.54}$Te$_{1.7}$Se$_{1.3}$. Because of the genuine bulk insulating behavior of these optimized BSTS crystals, the $\rho$-values obtained at low temperatures are not true resistivity values but depend on the thickness when $t \leq 1$~mm. An analysis of the resistance vs. thickness within a parallel resistor model of the resistance measured for crystals with different thicknesses shows that the surface contribution to the electrical transport amounts to 97~\% when the sample thickness is reduced to $1~\mu$m. Hence we conclude that devices fabricated with submicrometer thickness are sufficiently bulk insulating to exploit the topological surface states by transport techniques. This conclusion is supported by the observed collapse of the magnetoconductance data of an exfoliated BSTS nanoflake as a function of the perpendicular magnetic field component, further confirming 2D transport. The analysis within the HLN model for 2D weak antilocalization shows the fit parameter $\alpha \simeq -1$ as expected for conduction via a pair of topological surface states. Both this fact and our ARPES data recorded under band-bent conditions show a lack of Rashba-split non-topological surface states in our Bi$_{1.46}$Sb$_{0.54}$Te$_{1.7}$Se$_{1.3}$ crystals.

\ack{This work was part of the research program on Topological Insulators funded by FOM (Dutch Foundation for Fundamental Research of Matter). We thank V. Min for valuable contributions to earlier studies of the WAL effect. We acknowledge technical support from N. Plumb, N. Xu, M. Radovic and M. Shi during the ARPES measurements. EvH acknowledges support from the NWO Veni program. The research leading to these results has also received funding from the European Community's Seventh Framework Programme (FP7/2007-2013) under grant agreement No.312284 (CALIPSO).}

\section*{References}
\bibliography{RefsTI_NJP}

\providecommand{\newblock}{}
\begin{thebibliography}{10}
\expandafter\ifx\csname url\endcsname\relax
  \def\url#1{{\tt #1}}\fi
\expandafter\ifx\csname urlprefix\endcsname\relax\def\urlprefix{URL }\fi
\providecommand{\eprint}[2][]{\url{#2}}

\bibitem{Ren2011}
Ren Z, Taskin A~A, Sasaki S, Segawa K and Ando Y 2011 {\em Phys. Rev. B\/} {\bf
  84} 165311

\bibitem{Hasan&Kane2010}
Hasan M~Z and Kane C~L 2010 {\em Rev. Mod. Phys.\/} {\bf 82} 3045

\bibitem{Qi&Zhang2010}
Qi X~L and Zhang S~C 2011 {\em Rev. Mod. Phys.\/} {\bf 83} 1057

\bibitem{Zutic2004}
\v{Z}uti\'{c} I, Fabian J and Sarma S 2004 {\em Rev. Mod. Phys.\/} {\bf 76} 323

\bibitem{Qi2009}
Qi X~L, Hughes T~L, Raghu S and Zhang S~C 2009 {\em Phys. Rev. Lett.\/} {\bf
  102} 187001

\bibitem{Nayak2008}
Nayak C, Stern A, Freedman M and {Das Sarma} S 2008 {\em Rev. Mod. Phys.\/}
  {\bf 80} 1083

\bibitem{Hsieh2008}
Hsieh D, Qian D, Wray L, Xia Y, Hor Y~S, Cava R~J and Hasan M~Z 2008 {\em
  Nature\/} {\bf 452} 970

\bibitem{Xia2009}
Xia Y, Qian D, Hsieh D, Wray L, Pal A, Bansil A, Grauer D, Hor Y~S, Cava R~J
  and Hasan M~Z 2009 {\em Nature Phys.\/} {\bf 5} 398

\bibitem{Chen2009}
Chen Y~L, Analytis J~G, Chu J~H, Liu Z~K, Mo S~K, Qi X~L, Zhang H~J, Lu D~H,
  Dai X, Fang Z, Zhang S~C, Fisher I~R, Hussain Z and Shen Z~X 2009 {\em
  Science\/} {\bf 325} 178

\bibitem{Hsieh2009b}
Hsieh D, Xia Y, Qian D, Wray L, Meier F, Dil J~H, Osterwalder J, Patthey L,
  Fedorov A~V, Lin H, Bansil A, Grauer D, Hor Y~S, Cava R~J and Hasan M~Z 2009
  {\em Phys. Rev. Lett.\/} {\bf 103} 146401

\bibitem{Roushan2009}
Roushan P, Seo J, Parker C~V, Hor Y~S, Hsieh D, Qian D, Richardella A, Hasan
  M~Z, Cava R~J and Yazdani A 2009 {\em Nature\/} {\bf 460} 1106

\bibitem{Alpichshev2010}
Alpichshev Z, Analytis J~G, Chu J~H, Fisher I~R, Chen Y~L, Shen Z~X, Fang A and
  Kapitulnik A 2010 {\em Phys. Rev. Lett.\/} {\bf 104} 016401

\bibitem{Analytis2010}
Analytis J~G, McDonald R~D, Riggs S~C, Chu J~H, Boebinger G~S and Fisher I~R
  2010 {\em Nat. Phys.\/} {\bf 6} 960

\bibitem{Hor2009}
Hor Y~S, Richardella A, Roushan P, Xia Y, Checkelsky J~G, Yazdani A, Hasan M~Z,
  Ong N~P and Cava R~J 2009 {\em Phys. Rev. B\/} {\bf 79} 195208

\bibitem{Chen2010}
Chen J, Qin H~J, Yang F, Liu J, Guan T, Qu F~M, Zhang G~H, Shi J~R, Xie X~C,
  Yang C~L, Wu K~H, Li Y~Q and Lu L 2010 {\em Phys. Rev. Lett.\/} {\bf 105}
  176602

\bibitem{Checkelsky2011}
Checkelsky J~G, Hor Y~S, Cava R~J and Ong N~P 2011 {\em Phys. Rev. Lett.\/}
  {\bf 106} 196801

\bibitem{Ren2010}
Ren Z, Taskin A~A, Sasaki S, Segawa K and Ando Y 2010 {\em Phys. Rev. B\/} {\bf
  82} 241306(R)

\bibitem{Xiong2012}
Xiong J, Luo Y, Khoo Y~H, Jia S, Cava R~J and Ong N~P 2012 {\em Phys. Rev. B\/}
  {\bf 86} 045314

\bibitem{Jia2011}
Jia S, Ji H, Climent-Pascual E, Fuccillo M~K, Charles M~E, Xiong J, Ong N~P and
  Cava R~J 2011 {\em Phys. Rev. B\/} {\bf 84} 235206

\bibitem{Jia2012}
Jia S, Beidenkopf H, Drozdov I, Fuccillo M~K, Seo J, Xiong J, Ong N~P, Yazdani
  A and Cava R~J 2012 {\em Phys. Rev. B\/} {\bf 86} 165119

\bibitem{Ren2012}
Ren Z, Taskin A~A, Sasaki S, Segawa K and Ando Y 2012 {\em Phys. Rev. B\/} {\bf
  85} 155301

\bibitem{Fuccillo2013}
Fuccillo M, Jia S, Charles M and Cava R 2013 {\em J. Electr. Mater.\/} {\bf 42}
  1246

\bibitem{Taskin2011}
Taskin A~A, Ren Z, Sasaki S, Segawa K and Ando Y 2011 {\em Phys. Rev. Lett.\/}
  {\bf 107} 016801

\bibitem{Arakane2012}
Arakane T, Sato T, Souma S, Kosaka K, Nakyama N, Komatsu M, Takahashi T, Ren Z,
  Segawa K and Ando Y 2012 {\em Nat. Commun.\/} {\bf 3} 636

\bibitem{Tang2013}
Tang C~S, Xia B, Zou X, Chen S, Ou H~W, Wang L, Rusydi A, Zhu J~X and Chia
  E~E~M 2013 {\em Scientific Reports\/} {\bf 3} 3513

\bibitem{Ko2013}
Ko W, Jeon I, Kim H~W, Kwon H, Kahng S~J, Park J, Kim J~S, Hwang S~W and Suh H
  2013 {\em Scientific Reports\/} {\bf 3} 2656

\bibitem{Kim2014}
Kim S, Yoshizawa S, Ishida Y, Eto K, Segawa K, Ando Y, Shin S and Komori F 2014
  {\em Phys. Rev. Lett.\/} {\bf 112} 136802

\bibitem{Hikami1980}
Hikami S, Larkin A~I and Nagaoka Y 1980 {\em Prog. Theor. Phys. Rev.\/} {\bf
  63} 707

\bibitem{Lu&Shen2011}
Lu H~Z and Shen S~Q 2011 {\em Phys. Rev. B\/} {\bf 84} 125138

\bibitem{King2011}
King P~D~C, Hatch R~C, Bianchi M, Ovsyannikov R, Lupulescu C, Landolt G,
  Slomski B, Dil J~H, Guan D, Mi J~L, Rienks E~D~L, Fink J, Lindblad A,
  Svensson S, Bao S, Balakrishnan G, Iversen B~B, Osterwalder J, Eberhardt W,
  Baumberger F and Hofmann P 2011 {\em Phys. Rev. Lett.\/} {\bf 107} 096802

\bibitem{Bahramy2012}
Bahramy M~S, King P~D~C, de~la Torre A, Chang J, Shi M, Patthey L, Balakrishnan
  G, Hofmann P, Arita R, Nagaosa N and Baumberger F 2012 {\em Nature Commun.\/}
  {\bf 3} 1159

\bibitem{Xia2013}
Xia B, Ren P, Sulaev A, Liu P, Shen S~Q and Wang L 2013 {\em Phys. Rev. B\/}
  {\bf 87} 085442

\bibitem{Kim2014a}
Kim D~J, Xia J and Fisk Z 2014 {\em Nature Materials\/} {\bf 13} 466

\bibitem{Syers2014}
Syers P, Kim D, Fuhrer M~S and Paglione J 2014  (\textit{Preprint}
  \eprint{1408.3402})

\bibitem{Culcer2012}
Culcer D 2012 {\em Physica E\/} {\bf 44} 860

\bibitem{Lu&Shen2014}
Lu H~Z and Shen S~Q 2014 {\em Phys. Rev. Lett.\/} {\bf 112} 146601

\bibitem{Hsiung2013}
Hsiung T~S, Chen D~Y, Zhao L, Lin Y~H, Mou C~Y, Lee T~K, Wu M~K and Chen Y~Y
  2013 {\em Appl. Phys. Lett.\/} {\bf 103} 163111

\bibitem{Lee2012}
Lee J, Park J, Lee J~H, Kim J~S and Lee H~J 2012 {\em Phys. Rev. B\/} {\bf 86}
  245321

\end{thebibliography}

\end{document}